\documentclass[12 pt,twoside,prd,amsmath,amssymb,tightenlines,showpacs,
showkeys,checkins,eqsecnum]{revtex4}
\usepackage{epsfig}
\usepackage{psfig}
\usepackage{mathptmx}
\usepackage{bm}
\usepackage{graphicx}

\usepackage{hyperref}
\date{\filedate}

\renewcommand{\cal}{\mathcal}

\newcommand {\ve}{\varepsilon}

\def \myfigures #1#2#3#4#5#6#7#8
{\begin{figure}[ht]
    \begin{center}
        \includegraphics[width=#2 \textwidth]{#1.eps}
        \hfill
        \includegraphics[width=#6 \textwidth]{#5.eps}
        \parbox[t]{#4\textwidth}{\caption {#3}\label{#1}}
        \hfill
        \parbox[t]{#8\textwidth}{\caption {#7}\label{#5}}
    \end{center}
\end{figure}}

\def \myfigs #1#2#3#4#5#6#7#8
{\begin{figure}[ht]
    \begin{center}
        \includegraphics[width=#2 \textwidth, angle=-90]{#1.eps}
        \hfill
        \includegraphics[width=#6 \textwidth, angle=-90]{#5.eps}
        \vskip 5 mm
        \parbox[t]{#4\textwidth}{\caption {#3}\label{#1}}
        \hfill
        \parbox[t]{#8\textwidth}{\caption {#7}\label{#5}}
    \end{center}
\end{figure}}

\def\myfigure #1#2#3#4
{\begin{figure}[ht]\begin{center}
\includegraphics[width=#2 \textwidth, angle = -90]{#1.eps}
\parbox[t]{#4\textwidth}{\caption{#3}\label{#1}}
\end{center}\end{figure}}

\def\myfig #1#2#3#4
{\begin{figure}[ht]\begin{center}
\includegraphics[width=#2 \textwidth, angle = -90]{#1.eps}
\vskip 1 mm
\parbox[t]{#4\textwidth}{\caption{#3}\label{#1}}
\end{center}
\end{figure}}

\begin{document}
\title{Anisotropic cosmological models with perfect fluid and dark energy}
\author{Bijan \surname{Saha}}
\affiliation{Laboratory of Information Technologies\\
Joint Institute for Nuclear Research, Dubna\\
141980 Dubna, Moscow region, Russia} \email{saha@thsun1.jinr.ru, bijan@jinr.ru}
\homepage{http://thsun1.jinr.ru/~saha/}
\date{\today}
%\date{\filedate}
\begin{abstract}
We consider a self-consistent system of Bianchi type-I (BI)
gravitational field and a binary mixture of perfect fluid and dark
energy. The perfect fluid is taken to be the one obeying the usual
equation of state, i.e., $p = \zeta \ve$, with $\zeta \in [0,\,1]$
whereas, the dark energy density is considered to be either the
quintessence or the Chaplygin gas. Exact solutions to the
corresponding Einstein equations are obtained.
\end{abstract}



\maketitle

\bigskip

%%%%%%%%%%%%%%%%%%%%%%%%%%%%%%%%%%%%%%%%%%%%%%%%%%%%%%%%%%%%%%%%%%%%%%%%%%
            \section{Introduction}
%%%%%%%%%%%%%%%%%%%%%%%%%%%%%%%%%%%%%%%%%%%%%%%%%%%%%%%%%%%%%%%%%%%%%%%%%%

The description of the different phases of the Universe concerning
the time evolution of its acceleration field is among the main
objectives of the cosmological models. There is mounting evidence
that the Universe at present is dominated by the so-called dark
energy. Although the nature of the dark energy (DE) is currently
unknown, it is felt that DE is non-baryonic in origin
\cite{sahni}. It is also believed that the dark energy has large
negative pressure that leads to accelerated expansion of the
Universe.

In view of its importance in explaining the observational
cosmology many authors have considered cosmological models with
dark energy. The simplest example of dark energy is a cosmological
constant, introduced by Einstein in 1917 \cite{ein}. The discovery
that the expansion of the Universe is accelerating \cite{Bachall}
has promoted the search for new types of matter that can behave
like a cosmological constant \cite{caldwell,starobinsky} by
combining positive energy density and negative pressure. This type
of matter is often called {\it quintessence}. Zlatev {\it et al.}
\cite{zlatev} showed that "tracker field", a form of qiuntessence,
may explain the coincidence, adding new motivation for the
quintessence scenario.

An alternative model for the dark energy density was used by
Kamenshchik {\it et al.} \cite{kamen}, where the authors suggested
the use of some perfect fluid but obeying "exotic" equation of
state. This type of matter is known as {\it Chaplygin gas}. The
fate of density perturbations in a Universe dominated by the
Chaplygin gas, which exhibit negative pressure was studied by
Fabris {\it et al.} \cite{fabris}. Model with Chaplygin gas was
also studied in the Refs. \cite{dev,sen}. In a recent paper Kremer
\cite{kremer} has modelled the Universe as a binary mixture whose
constitutes are described by a van der Waals fluid and by a dark
energy density. In doing so the authors considered mainly a
spatially flat, homogeneous and isotropic Universe described by a
Friedmann-Robertson-Walker (FRW) metric.

The theoretical arguments and recent experimental data, which
support the existence of an anisotropic phase that approaches an
isotropic one, lead to consider the models of Universe with
anisotropic back-ground. The simplest of anisotropic models, which
nevertheless rather completely describe the anisotropic effects,
are Bianchi type-I (BI) homogeneous models whose spatial sections
are flat but the expansion or contraction rate is
direction-dependent. Moreover, a BI universe falls within the
general analysis of the singularity given by Belinskii et
al~\cite{belinskii} and evolves into a FRW universe~\cite{jacobs}
in  presence of a matter obeying the equation of state
$p\,=\,\zeta\,\ve, \quad \zeta < 1$. Since the modern-day Universe
is almost isotropic at large, this feature of the BI universe
makes it a prime candidate for studying the possible effects of an
anisotropy in the early Universe on present-day observations. In a
number of papers, e.g., \cite{PRD23501,PRD24010}, we have studied
the role of a nonlinear spinor and/or a scalar fields in the
formation of an anisotropic Universe free from initial
singularity. It was shown that for a suitable choice of
nonlinearity and the sign of $\Lambda$ term the model in question
allows regular solutions and the Universe becomes isotropic in the
process of evolution. Recently Khalatnikov {\it et al.}
\cite{khalat} studied the Einstein equations for a BI Universe in
the presence of dust, stiff matter and cosmological constant. In a
recent paper \cite{lambda} the author studied a self-consistent
system of Bianchi type-I (BI) gravitational field and a binary
mixture of perfect fluid and dark energy given by a cosmological
constant. The perfect fluid in that paper was chosen to be the one
obeying either the usual equation of state, i.e., $p = \zeta \ve$,
with $\zeta \in [0,\,1]$ or a van der Waals equation of state. In
this paper we study the evolution of an initially anisotropic
Universe given by a BI spacetime and a bimnary mixture of a
perfect fluid obeying the equation of state $p = \zeta \ve$ and a
dark energy given by either a quintessence or a Chaplygin gas.

%%%%%%%%%%%%%%%%%%%%%%%%%%%%%%%%%%%%%%%%%%%%%%%%%%%%%%%%%%%%%%%%%%%%%%%%%%
            \section{Basic equations}
%%%%%%%%%%%%%%%%%%%%%%%%%%%%%%%%%%%%%%%%%%%%%%%%%%%%%%%%%%%%%%%%%%%%%%%%%%
The gravitational field in our
case is given by a Bianchi type I (BI) metric in the form
\begin{equation}
ds^2 =  dt^2 - a^2 dx^2 - b^2 dy^2 - c^2 dz^2,
\label{BI}
\end{equation}
with the metric functions $a,\,b,\,c$ being the functions of time
$t$ only.

The Einstein field equations for the BI space-time we write in the form
\begin{subequations}
\label{ee}
\begin{eqnarray}
\frac{\ddot b}{b} +\frac{\ddot c}{c} + \frac{\dot b}{b}\frac{\dot
c}{c}&=&  \kappa T_{1}^{1},\label{11}\\
\frac{\ddot c}{c} +\frac{\ddot a}{a} + \frac{\dot c}{c}\frac{\dot
a}{a}&=&  \kappa T_{2}^{2},\label{22}\\
\frac{\ddot a}{a} +\frac{\ddot b}{b} + \frac{\dot a}{a}\frac{\dot
b}{b}&=&  \kappa T_{3}^{3},\label{33}\\
\frac{\dot a}{a}\frac{\dot b}{b} +\frac{\dot b}{b}\frac{\dot c}{c}
+\frac{\dot c}{c}\frac{\dot a}{a}&=&  \kappa T_{0}^{0}.
\label{00}
\end{eqnarray}
\end{subequations}
Here $\kappa$ is the Einstein gravitational constant and over-dot means
differentiation with respect to $t$. The energy-momentum tensor of the
source is given by
\begin{equation}
T_{\mu}^{\nu} = (\ve + p) u_\mu u^\nu - p \delta_\mu^\nu,
\label{emt}
\end{equation}
where $u^\mu$ is the flow vector satisfying
\begin{equation}
g_{\mu\nu} u^\mu u^\nu = 1.
\label{scprod}
\end{equation}
Here $\ve$ is the total energy density of a perfect fluid and/or
dark energy density, while $p$ is the corresponding pressure. $p$
and $\ve$ are related by an equation of state which will be
studied below in detail. In a co-moving system of coordinates from
\eqref{emt} one finds
\begin{equation}
T_0^0 = \ve, \qquad T_1^1 = T_2^2 = T_3^3 = - p.
\label{compemt}
\end{equation}
In view of \eqref{compemt} from \eqref{ee} one immediately obtains
\cite{PRD23501}
\begin{subequations}
\label{abc}
\begin{eqnarray}
a(t) &=&
D_{1} \tau^{1/3} \exp \bigl[X_1 \int\,\frac{dt}{\tau (t)} \bigr],
\label{a} \\
b(t) &=&
D_{2} \tau^{1/3} \exp \bigl[X_2 \int\,\frac{dt}{\tau (t)} \bigr],
\label{b}\\
c(t) &=&
D_{3} \tau^{1/3}\exp \bigl[X_3  \int\,\frac{dt}{\tau (t)} \bigr].
\label{c}
\end{eqnarray}
\end{subequations}
Here $D_i$ and $X_i$ are some arbitrary constants obeying
$$D_1 D_2 D_3 = 1, \qquad X_1 + X_2 + X_3 = 0,$$
and $\tau$ is a function of $t$ defined to be
\begin{equation}
\tau = a b c. \label{tau}
\end{equation}
From \eqref{ee} for $\tau$ one find
\begin{equation}
\frac{\ddot \tau}{\tau} = \frac{3 \kappa}{2} \bigl(\ve - p\bigr).
\label{dtau}
\end{equation}
On the other hand the conservation law for the energy-momentum tensor gives
\begin{equation}
\dot{\ve} = -\frac{\dot \tau}{\tau} \bigl(\ve + p\bigr).
\label{dve}
\end{equation}
After a little manipulations from \eqref{dtau} and \eqref{dve} we
find
\begin{equation}
\dot \tau = \pm \sqrt{C_1 + 3 \kappa \ve \tau^2}, \label{1st}
\end{equation}
with $C_1$ being an integration constant. On the other hand
rewriting \eqref{dve} in the form
\begin{equation}
\frac{\dot{\ve}}{(\ve + p)} = -\frac{\dot \tau}{\tau},
\label{dve1}
\end{equation}
and taking into account that the pressure and the energy density
obey a equation of state of type $p = f (\ve)$, we conclude that
$\ve$ and $p$, hence the right hand side of the Eq. \eqref{dtau}
is a function of $\tau$ only, i.e.,
\begin{equation}
\ddot \tau = \frac{3 \kappa}{2} \bigl(\ve - p\bigr)\tau \equiv
{\cal F} (\tau). \label{dtau1}
\end{equation}
From the mechanical point of view Eq. \eqref{dtau1} can be
interpreted as an equation of motion of a single particle with
unit mass under the force ${\cal F}(\tau)$. Then the following
first integral exists \cite{landau}:
\begin{equation}
\dot \tau = \sqrt{2[{\cal E} - {\cal U} (\tau)]}. \label{first}
\end{equation}
Here ${\cal E}$ can be viewed as energy and ${\cal U}(\tau)$ is
the potential of the force ${\cal F}$. Comparing the Eqs.
\eqref{1st} and \eqref{first} one finds ${\cal E} = C_1 /2$ and
\begin{equation}
{\cal U} (\tau) = - \frac{3}{2} \kappa \ve \tau^2.
\label{potential}
\end{equation}

Finally, rearranging \eqref{1st}, we write the solution to the Eq.
\eqref{dtau} in quadrature:
\begin{equation}
\int \frac{d\tau}{\sqrt{C_1 + 3\kappa \ve \tau^2}} = t + t_0,
\label{quad}
\end{equation}
where the integration constant $t_0$ can be taken to be zero,
since it only gives a shift in time.

In what follows we study the Eqs. \eqref{dtau} and \eqref{dve} for
perfect fluid and/or dark energy for different equations of state
obeyed by the source fields.

%%%%%%%%%%%%%%%%%%%%%%%%%%%%%%%%%%%%%%%%%%%%%%%%%%%%%%%%%%%%%%%%%%
\section{Universe as a binary mixture of perfect fluid and dark energy}
%%%%%%%%%%%%%%%%%%%%%%%%%%%%%%%%%%%%%%%%%%%%%%%%%%%%%%%%%%%%%%%%%%

In this section we thoroughly study the evolution of the BI
Universe filled with perfect fluid and dark energy in details.
Taking into account that the energy density ($\ve$) and pressure
($p$) in this case comprise those of perfect fluid and dark
energy, i.e.,
$$\ve = \ve_{\rm pf} + \ve_{\rm DE}, \qquad p = p_{\rm pf} + p_{\rm
DE}$$ the energy momentum tensor can be decomposed as
\begin{equation}
T_{\mu}^{\nu} = (\ve_{\rm DE} + \ve_{\rm pf} + p_{\rm DE} + p_{\rm
pf} ) u_\mu u^\nu - (p_{\rm DE} + p_{\rm pf}) \delta_\mu^\nu.
\label{emtde}
\end{equation}
In the above equation $\ve_{\rm DE}$ is the dark energy density,
$p_{\rm DE}$ its pressure. We also use the notations $\ve_{\rm
pf}$ and $p_{\rm pf}$ to denote the energy density and the
pressure of the perfect fluid, respectively. Here we consider the
case when the perfect fluid in question obeys the following
equation of state
\begin{equation}
p_{\rm pf}\,=\,\zeta\,\ve_{\rm pf}. \label{eqst}
\end{equation}
Here $\zeta$ is a constant and lies in the interval $\zeta\, \in
[0,\,1]$. Depending on its numerical value, $\zeta$ describes the
following types of Universes \cite{jacobs}
\begin{subequations}
\label{zeta}
\begin{eqnarray}
\zeta &=& 0, \qquad \qquad {\rm (dust\,\, Universe)},\\
\zeta &=& 1/3, \quad \qquad {\rm (radiation\,\, Universe)},\\
\zeta &\in& (1/3,\,1), \quad {\rm (hard\,\, Universes)},\\
\zeta &=& 1, \quad \qquad \quad {\rm (Zel'dovich\,\, Universe \,\,
or\,\, stiff\,\, matter)}.
\end{eqnarray}
\end{subequations}

In a comoving  frame the conservation law of the energy momentum
tensor leads to the balance equation for the energy density
\begin{equation}
\dot{\ve}_{\rm DE} + \dot{\ve}_{\rm pf} = - \frac{\dot \tau}{\tau}
\bigl(\ve_{\rm DE} + \ve_{\rm pf} + p_{\rm DE} + p_{\rm pf}
\bigr). \label{dveden}
\end{equation}
The dark energy is supposed to interact with itself only and it is
minimally coupled to the gravitational field. As a result the
evolution equation for the energy density decouples from that of
the perfect fluid, and from  Eq. \eqref{dveden} we obtain two
balance equations
\begin{subequations}
\label{balance}
\begin{eqnarray}
\dot{\ve}_{\rm DE} + \frac{\dot \tau}{\tau} \bigl(\ve_{\rm DE} +
p_{\rm DE}\bigr) &=& 0,
\label{deve}\\
\dot{\ve}_{\rm pf} + \frac{\dot \tau}{\tau} \bigl(\ve_{\rm pf} +
p_{\rm pf}\bigr) &=& 0. \label{pfve}
\end{eqnarray}
\end{subequations}
In view of the Eq. \eqref{eqst} from \eqref{pfve} one easily finds
\begin{equation}
\ve_{\rm pf} = \ve_0/\tau^{(1+\zeta)}, \qquad p_{\rm pf} = \ve_0
\zeta/\tau^{(1+\zeta)},
 \label{vepf}
\end{equation}
where $\ve_0$ is the integration constants. In absence of the dark
energy one immediately finds
\begin{equation}
\tau = C t^{2/(1+\zeta)}, \label{tpf}
\end{equation}
with $C$ being some integration constant. As one sees from
\eqref{abc}, in absence of a $\Lambda$ term, for $\zeta < 1$ the
initially anisotropic Universe eventually evolves into an
isotropic FRW one, whereas, for $\zeta = 1$, i.e., in case of
stiff matter the isotropization does not take place.

In what follows we consider the case when the Universe is filled
with the dark energy as well.

\myfigs{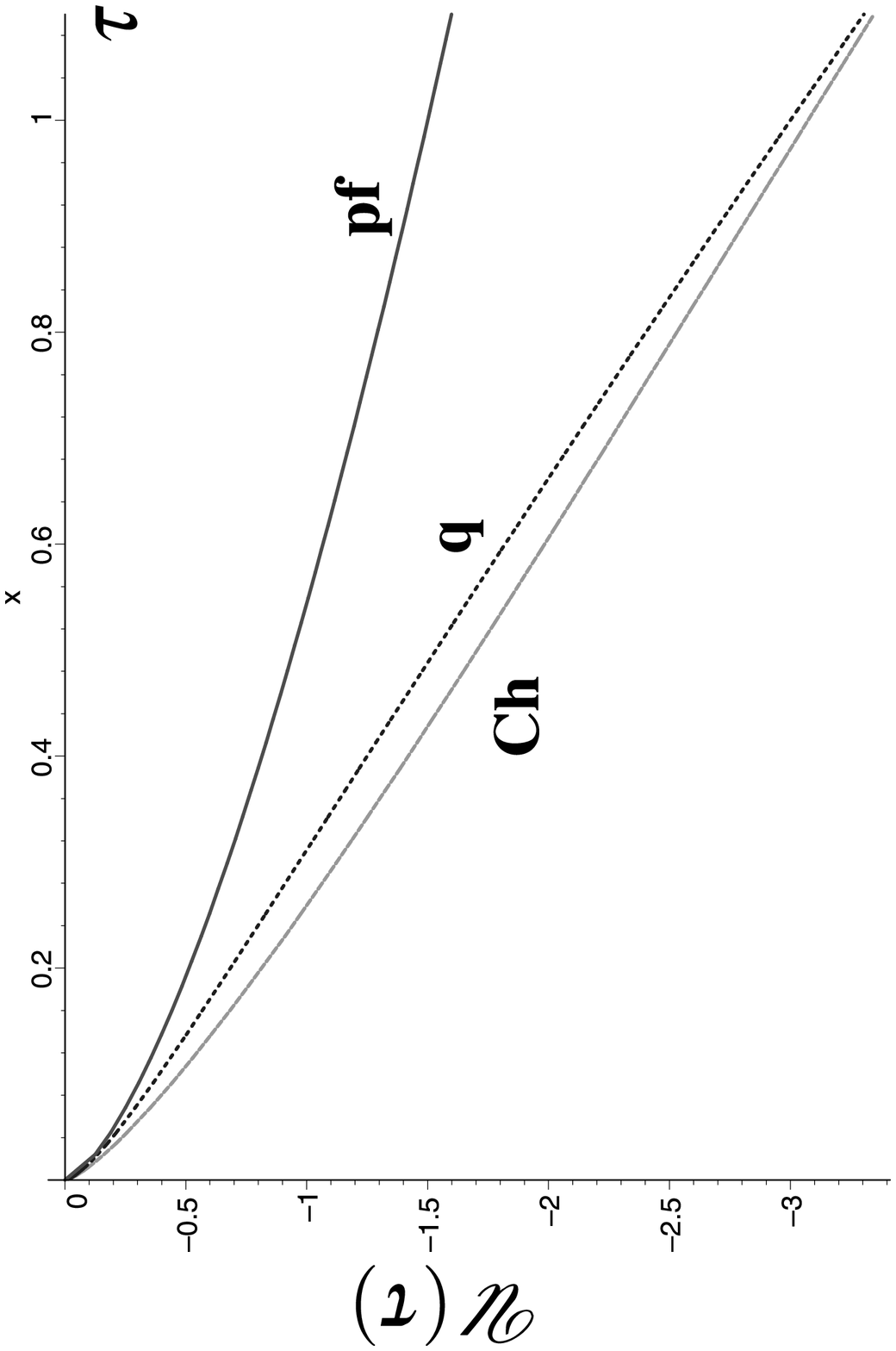}{0.30}{View of potentials when the Universe is
filled with perfect fluid, perfect fluid plus quintessence and
perfect fluid plus Chaplygin gas,
respectively.}{0.43}{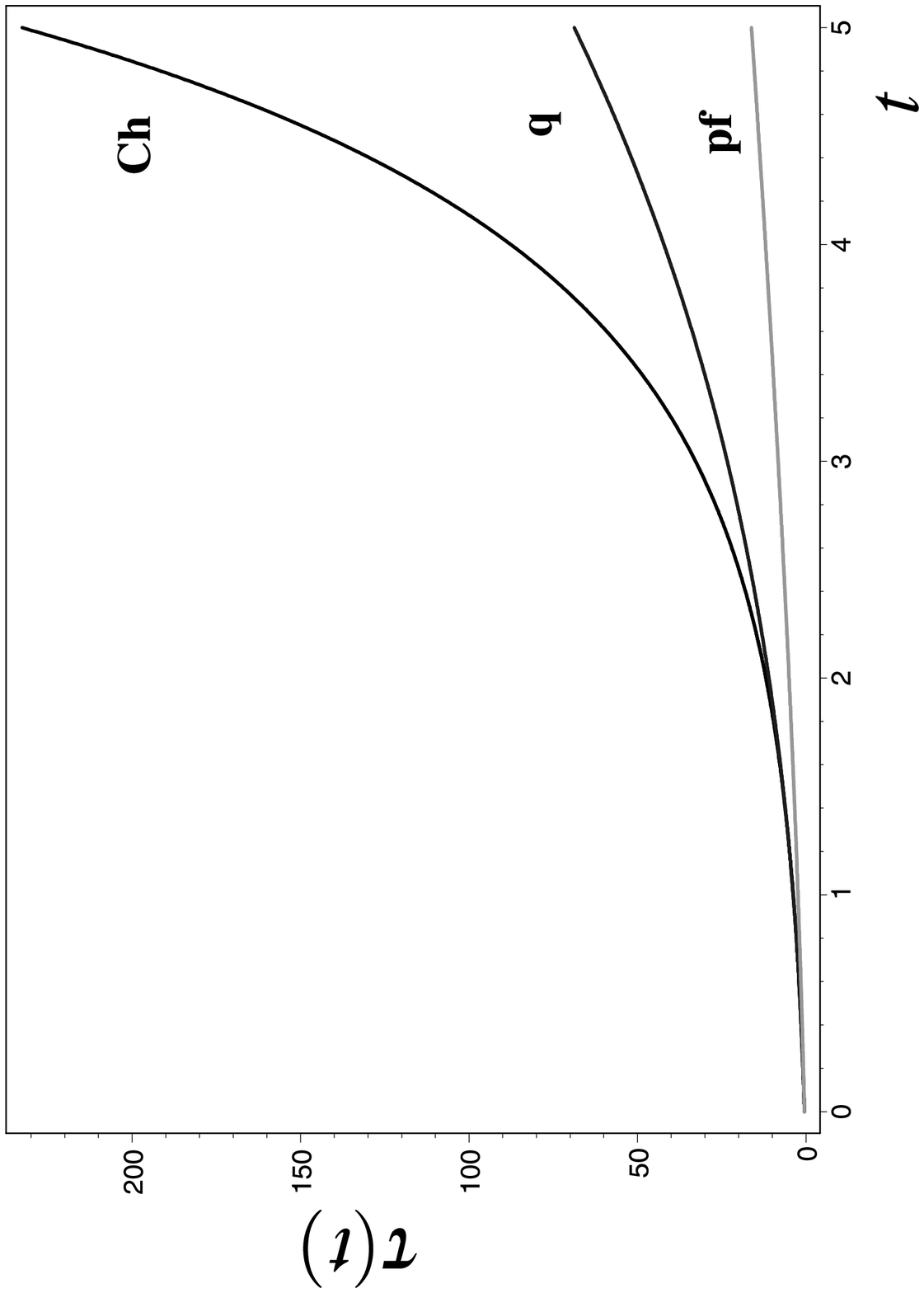}{0.30}{Evolution of the BI Universe
corresponding to the potentials illustrated in Fig.
\ref{poten}.}{0.43}

In Fig. \ref{poten} we have plotted the potentials when the
Universe is filled with perfect fluid, perfect fluid plus
quintessence and perfect fluid plus Chaplygin gas, respectively.
The perfect fluid is given by a radiation. As one sees, these
types of potentials allows only infinite motion, i.e., the
Universe expands infinitely. The Fig. \ref{taupqc} shows the
evolution of the BI Universe. The introduction of dark energy
results in accelerated expansion of the Universe. The view of
acceleration has been illustrated in Fig. \ref{acpqc}.

\myfigs{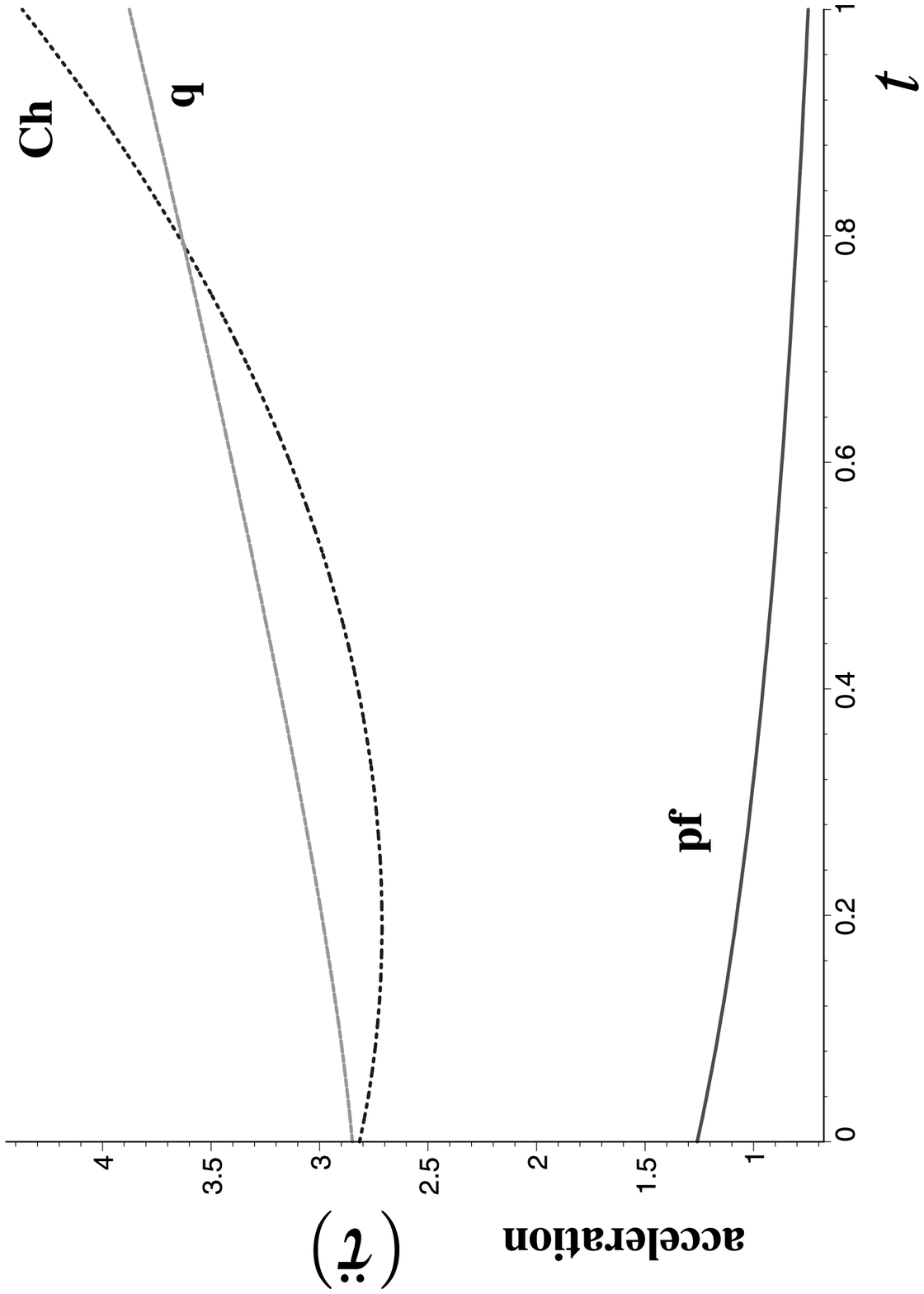}{0.30}{Acceleration of a BI Universe filled with a
perfect fluid, perfect fluid plus quintessence and perfect fluid
plus Chaplygin gas, respectively.}{0.43}{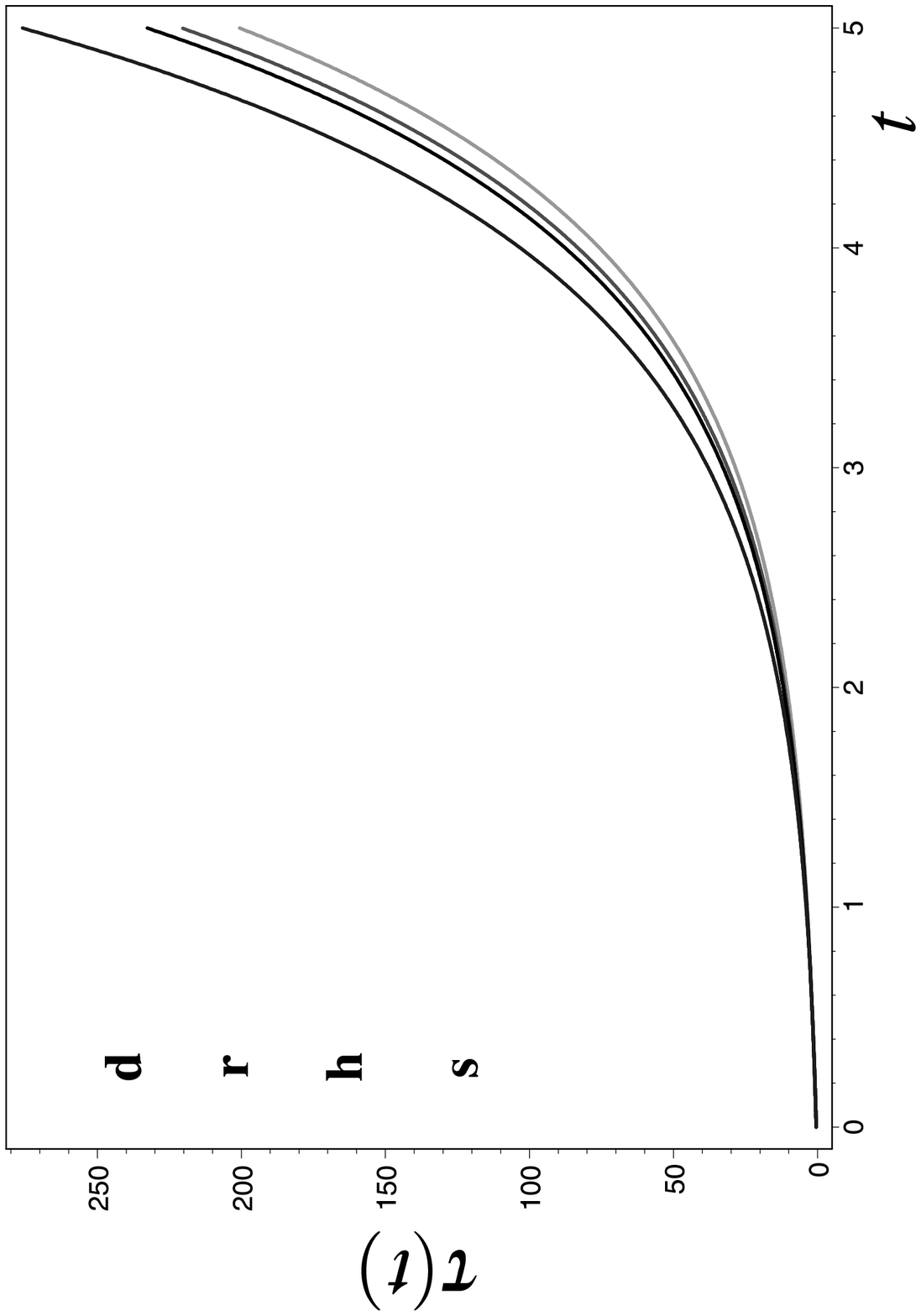}{0.30}{Evolution of
the BI Universe filled with a perfect fluid and Chaplygin gas.
}{0.43}

The Fig. \ref{tauc} shows the evolution of a BI Universe filled
with a perfect fluid and a Chaplygin gas. Here {\bf "d", "r", "h"}
and {\bf "s"} stand for dust, radiation, hard Universe and stiff
matter, respectively. As one sees, even in case of a stiff matter
the Universe expands rapid enough to evolve into an isotropic one.

\subsection{Case with a quintessence}

Let us consider the case when the dark energy is given by a
quintessence. As it was mentioned earlier, a new type of matter,
often known as quintessence, can behave like a cosmological
constant and was constructed by combining positive energy density
and negative pressure and obeys the equation of state
\begin{equation}
p_{\rm q} = w_{\rm q} \ve_{\rm q}, \label{quint}
\end{equation}
where the constant $w_{\rm q}$ varies between $-1$ and zero, i.e.,
$w_{\rm q} \in [-1,\,0]$. In account of \eqref{quint} from
\eqref{deve} one finds
\begin{equation}
\ve_{\rm q} = \ve_{0 \rm q}/\tau^{(1+w_{\rm q})}, \qquad p_{\rm q}
= w_{\rm q}\ve_{0 \rm q}/\tau^{(1+w_{\rm q})}, \label{veq}
\end{equation}
with $\ve_{0 \rm q}$ being some integration constant.

Now the evolution equation for $\tau$ \eqref{dtau} can be written
as
\begin{equation}
\ddot \tau = \frac{3 \kappa}{2} \Bigl(
\frac{(1-\zeta)\ve_0}{\tau^\zeta} + \frac{(1- w_{\rm q})\ve_{0 \rm
q}}{\tau^{w_{\rm q}}}\Bigr). \label{dtauq}
\end{equation}
As it was mentioned earlier the Eq. \eqref{dtauq} admits exact
solution that can be written in quadrature as
\begin{equation}
\int \frac{d\tau}{\sqrt{C_1 + 3 \kappa\bigl(\ve_0 \tau^{(1
-\zeta)} + \ve_{0 \rm q}\tau^{(1 -w_{\rm q})} \bigr)}} = t + t_0.
\label{quadq}
\end{equation}
Here $t_0$ is a constant of integration that can be taken to be
trivial.

\myfigs{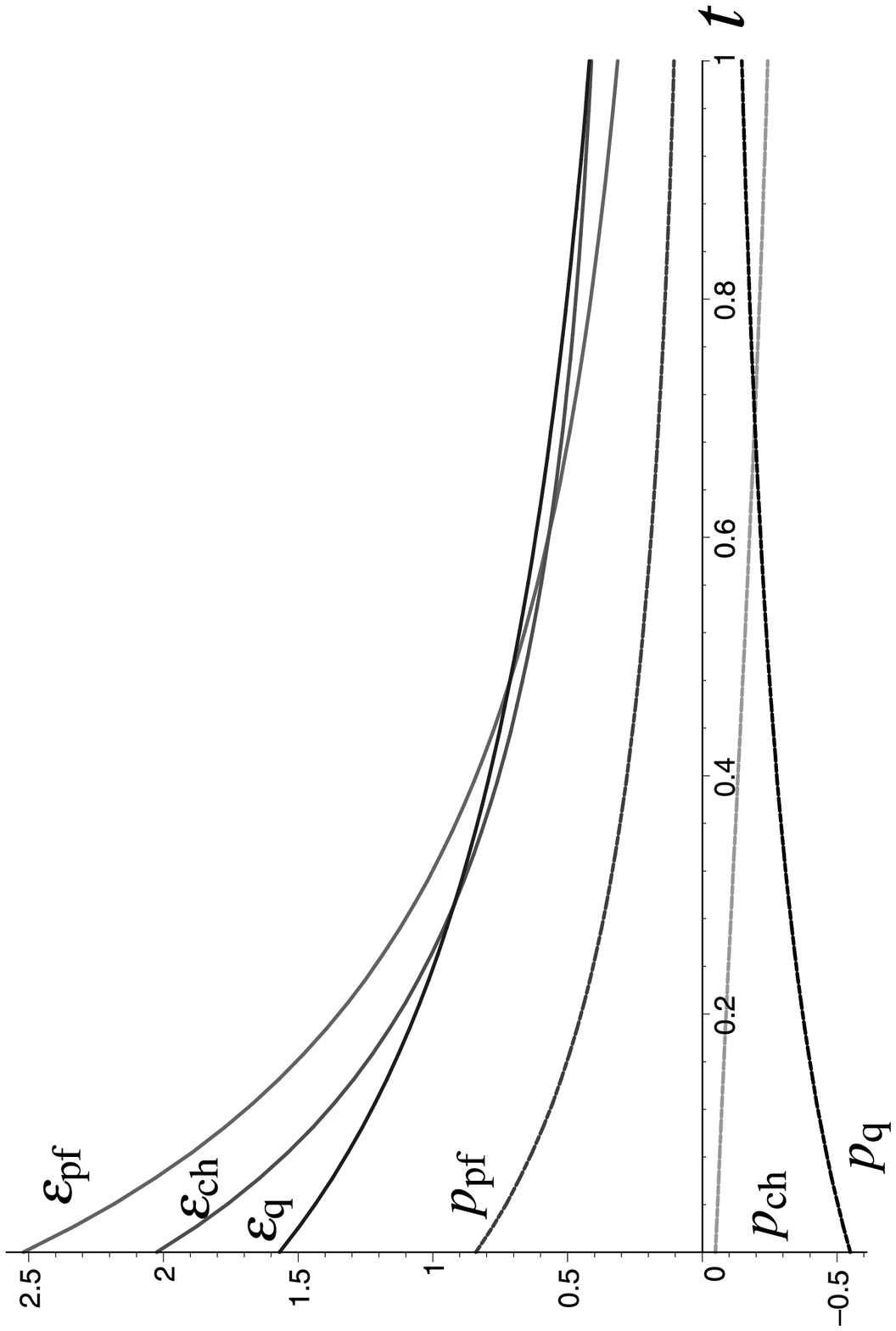}{0.30}{View of energy density and corresponding
pressure when the Universe is filled with a perfect fluid,
quintessence and Chaplygin gas,
respectively.}{0.43}{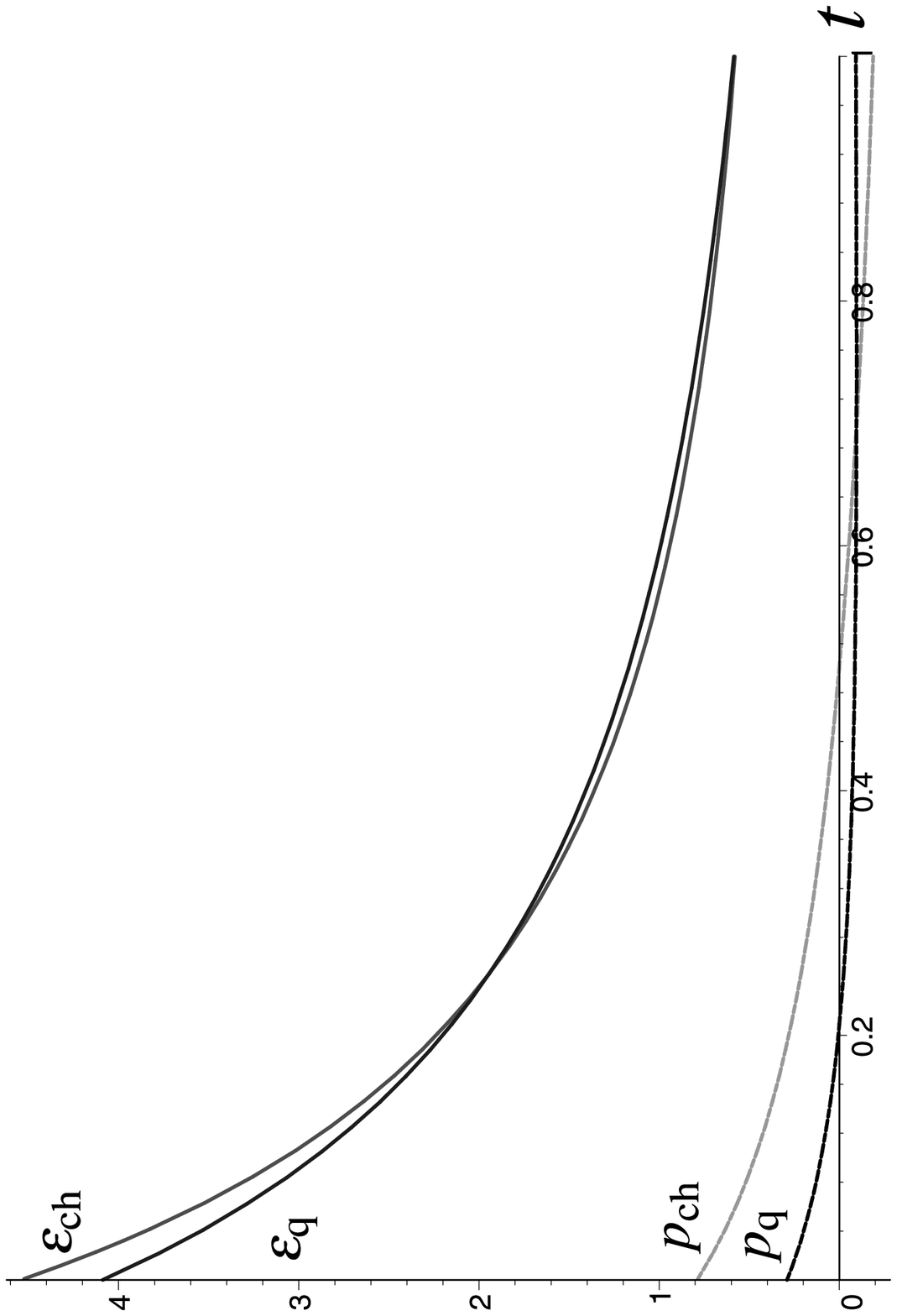}{0.30}{View of energy density and
corresponding pressure when the Universe is a binary mixture of a
perfect fluid and quintessence and a perfect fluid and Chaplygin
gas, respectively.}{0.43}

\subsection{Case with Chaplygin gas}

Let us now consider the case when the dark energy is represented
by a Chaplygin gas. We have already mentioned that the Chaplygin
gas was suggested as an alternative model of dark energy with some
exotic equation of state, namely
\begin{equation}
p_{\rm c} = -A/\ve_{\rm c}, \label{chaplyn}
\end{equation}
with $A$ being a positive constant. In view of the Eq.
\eqref{chaplyn} from \eqref{deve} one now obtains
\begin{equation}
\ve_{\rm c} = \sqrt{\ve_{0 \rm c}/\tau^2 + A}, \qquad p_{\rm c} =
-A/\sqrt{\ve_{0 \rm c}/\tau^2 + A}, \label{vech}
\end{equation}
with $\ve_{0 \rm c}$ being some integration constant.

Proceeding analogously as in previous case for $\tau$ we now have
\begin{equation}
\ddot \tau = \frac{3 \kappa}{2} \Bigl(
\frac{(1-\zeta)\ve_0}{\tau^\zeta} + \sqrt{\ve_{0 \rm c} + A
\tau^2} + A/\sqrt{\ve_{0 \rm c} + A \tau^2}\Bigr). \label{dtauc}
\end{equation}
The corresponding solution in quadrature now has the forms:
\begin{equation}
\int \frac{d\tau}{\sqrt{C_1 + 3 \kappa\bigl(\ve_0 \tau^{(1
-\zeta)} + \sqrt{\ve_{0 \rm c}\tau^2 + A \tau^4}\bigr)}} = t,
\label{quadc}
\end{equation}
where the second integration constant has been taken to be zero.

\section{Conclusion}

A self-consistent system of BI gravitational field filled with a
perfect fluid and a dark energy has been considered. The exact
solutions to the corresponding field equations are obtained. The
inclusion of the dark energy into the system gives rise to an
accelerated expansion of the model. As a result the initial
anisotropy of the model quickly dies away. Note that the
introduction of the dark energy does not eliminate the initial
singularity.

\newcommand{\hnl}{\htmladdnormallink}

\end{document}